\documentclass[a4paper,11pt]{article}
\pdfoutput=1 

\usepackage{jcappub} 

\usepackage[T1]{fontenc} 

\usepackage{caption}
\usepackage{subcaption}
\usepackage{graphicx}
\usepackage{mathrsfs}
\usepackage{amsmath}
\usepackage{amssymb}
\usepackage{mathtools} 
\usepackage{comment}

\providecommand{\abs}[1]{\lvert#1\rvert}

\providecommand{\ro}[1]{\mathrm{#1}}

\providecommand{\Mp}{M_{\mathrm{Pl}}}

\title{Noninflationary solution to the monopole problem}

\author[]{Daniele Perri}

\affiliation[]{Institute of Theoretical Physics, Faculty of Physics, University of
Warsaw, \\ ul. Pasteura 5, PL-02-093 Warsaw, Poland\\}

\emailAdd{daniele.perri@fuw.edu.pl}

\abstract{
Magnetic monopoles are a long-standing prediction of Grand Unified Theories, yet their efficient production in early universe phase transitions would lead to a monopole abundance that far exceeds observational limits. The standard solution of the problem invokes inflation occurring after monopole production, diluting their density to undetectable levels and eliminating any possibility of present-day observation. 
Here, we propose an alternative solution based on the breaking, in the early universe prior to Big Bang Nucleosynthesis, of the Weyl conformal symmetry of the gauge kinetic sector of the Lagrangian. This mechanism enhances monopole annihilation, thereby reducing their abundance to acceptable levels without requiring inflation. This scenario also predicts a residual flux of GUT monopoles potentially within the sensitivity of current and upcoming cosmic ray detectors, making their discovery possible in the near future.
}

\begin{document}
\maketitle
\flushbottom

\section{Introduction}
\label{sec:intro}

The quest for magnetic monopoles has now a century of history, starting with the first model proposed by Dirac \cite{Dirac:1931kp}. Such particles would be crucial for the symmetrization of the Maxwell equations, giving also a theoretical explanation for the quantization of the electric charge. More recently, models of magnetic monopoles have been proposed as topological defects of the vacuum manifold in gauge theories with nontrivial second homotopy group. These monopoles are usually called 't Hooft-Polyakov monopoles \cite{tHooft:1974kcl, Polyakov:1974ek}, and can be produced during phase transition in the early universe \cite{Preskill:1979zi,Zeldovich:1978wj,Vilenkin:2000jqa}. A Significant example of 't Hooft-Polyakov monopoles is the one associated to the symmetry breaking of Grand Unified Theories (GUTs), whose lagrangians, in most models, admit monopole solutions \cite{Goddard_1978,PhysRevD.21.2940,PhysRevD.27.2119}.

The abundance of 't Hooft-Polyakov monopoles produced during a phase transition are usually computed via the Kibble mechanism \cite{Kibble:1976sj}, based on the approximation of one monopole produced in each Hubble volume at the time of the phase transition. Such an estimate provides an overabundant amount of monopoles that easily overclose the universe if the critical temperature of the phase transition is $T_c \gtrsim 10^{11}~\mathrm{GeV}$ \cite{Kolb:1990vq}. This result is often addressed in the literature as the ``monopole problem'', because the production of monopoles at the GUT scale would be in contradiction with the energy density of the universe.

Several potential solutions to the monopole problem have been proposed.  
Currently, the most common explanation is the inflationary Universe model. The rapid expansion that comes with inflation allows a small, sub-horizon region of space where the Higgs field is coherent to extend across the entire observable Universe. 
As a consequence, there would be less than one monopole in the observable universe. 
At the time when the first models of inflation were proposed, the monopole problem was considered so compelling that solving it was one of the strongest elements in support of the inflationary hypothesis \cite{Linde:1981mu}. 
Alternative solutions based on more complex patterns of symmetry breaking have been also proposed in the literature. Among these, we mention here two possibilities. 
First, if there is no complete unification of the forces (for example, if the original symmetry group includes $U(1)$ symmetries, as in $G = H \otimes U(1)$) or if the GUT symmetry is not fully restored in the early Universe (e.g., if the maximum temperature reached by the Universe was below $T_c$), then the monopole problem would not arise, since monopoles would have never been produced.
A second solution proposed in \cite{Langacker:1980kd} is based on the unusual symmetry breaking pattern $SU(5) \to SU(3) \otimes SU(2) \otimes U(1) \to SU(3) \to SU(3) \otimes U(1)$.
In this case, the monopoles are produced during the first phase transition, but between the second and the last phase transition, when the $U(1)$ of the electromagnetism is spontaneously broken, the Universe transitions to a superconducting state. Consequently, the magnetic flux confines into flux tubes, enhancing the annihilation rate of monopoles and antimonopoles. 
In this scenario, the final monopole density is approximately one per horizon volume by the conclusion of the superconducting phase, and is therefore smaller than in standard symmetry breaking patterns.
Alternative solutions to the monopole problem have also been proposed in \cite{Stojkovic:2004hz, Stojkovic:2005zh}.

In this work, we propose a solution to the monopole problem, alternative to the standard inflationary paradigm, that does not involve any modification of the gauge symmetries of the lagrangian, but instead of the kinetic term of the gauge field, breaking the Weyl conformal symmetry of the gauge sector.
As in the case of \cite{Langacker:1980kd}, we then provide a mechanism to reduce the abundance of magnetic monopoles after the time of production by enhancing the process of monopole-antimonopole annihilation.
In particular, our model predicts that monopoles would firstly be produced in the early Universe as global defects \cite{PhysRevLett.63.341, Linde:1990flp}, and only later would acquire a charge under the electromagnetic interaction.
The very efficient mechanism of global monopole annihilation reduces the monopole abundance to $\sim 4$ monopoles per Hubble volume \cite{PhysRevLett.65.1709}, drastically reducing the number of monopoles at the time when they acquire a magnetic charge.
We demonstrate that this mechanism can solve the monopole problem even for GUT-scale monopoles produced after the end of inflation, providing a viable mechanism for monopole production.
Similar modifications to the gauge-field kinetic term have been explored in the context of primordial magnetogenesis \cite{Turner:1987bw,Kunze_2010,Kunze:2012rq,Ratra:1991bn,Demozzi:2009fu,Ferreira:2014hma,Kobayashi:2014sga,Bamba:2003av,Martin:2007ue,Gasperini:1995dh}.
Here, however, we are only interested in the consequences for the abundance of monopoles.
In previous studies, such modifications have been motivated either by non-minimal couplings to a time-dependent gravitational background \cite{Turner:1987bw,Kunze_2010,Kunze:2012rq}, or by couplings to time-dependent scalar field, such as oscillating inflaton fields, dilaton fields, or other background scalars \cite{Ratra:1991bn, Martin:2007ue, Demozzi:2009fu, Gasperini:1995dh}. Dilatonic couplings, in particular, arise naturally in UV completions inspired by higher-dimensional string theories.

This paper is organized as follows. In Section \ref{sec:global_iff}, we review the theory of global monopoles and models of minimally coupled $I^2 FF$ theories, which break the conformal symmetry of the gauge sector. 
In Section \ref{sec:solution}, we show how such theories provide a solution to the monopole problem without the need for an inflationary epoch after the monopole production.
We then conclude in Section \ref{sec:conclusion}. In Appendix \ref{sec:phase_transition}, we review the standard mechanism of magnetic monopole production during phase transitions and we compute the predicted abundance, and in Appendix \ref{sec:dilatonic}, we discuss the case of a dilatonic coupling of the $I^2$ function to the lagrangian of the model.
Throughout this work, we fix the spacetime metric to a flat FRW, and we choose the metric tensor signature $(+---)$.
We adopt Heaviside-Lorentz units, with $c = \hbar = k_B = 1$, and use $\Mp$ to denote the reduced Planck mass~$ (8 \pi G)^{-1/2}$.
We use Greek letters for spacetime indices and Latin letters for internal quantum numbers.

\section[Global monopoles and $I^2 FF$ theories]{Global monopoles and $\mathbf{I^2 FF}$ theories}
\label{sec:global_iff}

In this section, we review the main features of the global monopole solution. We then show how an effective global monopole solution can be defined in theories that breaks the conformal invariance of the lagrangian kinetic gauge sector.

\subsection{The global monopole solution}

In the case of the spontaneous breaking Of a global symmetry, the lagrangian may still admit monopole solutions. These monopoles do not carry magnetic charge but instead posses a global charge associated with the unbroken global symmetry group. They are commonly referred to as ``global monopoles'', in contrast to the standard gauge monopoles, which we here address also as ``local''. Global monopoles exhibit distinctive and exotic properties, which we review in detail below.

Analogously to the case of 't Hooft-Polyakov monopoles, global monopoles are produced during the spontaneous symmetry breaking of a global symmetry $G$ into a smaller group $H$ such that the second homotopy group of the vacuum manifold $G/H$ is nontrivial, i.e., $\Pi_2 \left( G/H \right) \neq I$.
The simplest model where a global monopole solution can be defined is the global version of the Georgi-Glashow model:
\begin{equation}
    \frac{\mathcal{L}}{\sqrt{-g}}  = - \frac{1}{2} (\partial_\mu \phi)^a (\partial_\mu \phi)^a - \frac{\lambda}{4} (\phi^a \phi^a - v^2)^2 .
\end{equation}
Here $\phi^{\rm a}$ is a triplet of $SO(3)$ and the vacuum expectation value $v$ breaks the $SO(3)$ global symmetry into a $SO(2)$ global symmetry.

The global monopole solution is characterized by the topological charge \cite{Vilenkin:2000jqa}
\begin{equation}
    N = \frac{1}{8 \pi} \oint dS^{ij} \abs{\phi}^{-3} \epsilon_{a b c} \phi^a \partial_i \phi^b \partial_j \phi^c ,
\end{equation}
which corresponds to the global charge of the monopole.
In the fundamental case $N=1$, the solution is the spherically-symmetric configuration:
\begin{equation}
    \phi^{a} = v h(r) \frac{x^{a}}{r} ,
\end{equation}
where the function $h(r)$ vanishes for $r = 0$ and approaches $1$ in the limit $r \rightarrow \infty$.\footnote{Note that in this case the symmetry is global, and therefore no monopole solution exists for an associated vector boson field, unlike in the gauge case.} In the following of this work, we always assume the minimal case $N=1$.

As in the gauge case, one can compute the total energy of the global monopole by integrating the time-time component of the stress-energy tensor outside the monopole core of size $\delta \sim (\sqrt{\lambda} v)^{-1}$ \cite{Vilenkin:2000jqa},
\begin{equation}
    E \sim 4 \pi \int^{R}_{\delta} T^t_t r^2 dr \sim 4 \pi v^2 R,
\end{equation}
where $R$ is a cut-off radius. This integral is clearly divergent for $R \to \infty$. However, for practical realizations, a cutoff radius can be approximately fixed to the distance of the closest antimonopole. 
So, although one isolated global monopole would have infinite energy, the problem is solved in a physical realization of the symmetry breaking, where the coherence length of the Higgs field is finite.

As the monopole total energy depends on $R$ linearly, the attractive force between a monopole and an antimonopole is independent of distance,
\begin{equation}
    F = \frac{\partial E}{\partial R} \sim 4 \pi v^2 .
\end{equation}
The strong attractive force between global monopole–antimonopole pairs makes their annihilation highly efficient. Numerical simulations \cite{PhysRevLett.65.1709} indicate that the system rapidly evolves into a scale-invariant regime in less than a Hubble time, with a monopole number density
\begin{equation}
\label{eq:nGlobal}
    n_{\rm M} = \left(4 \pm 1.5 \right) R_{\mathrm{H}}^{-3} ,
\end{equation}
where $R_{\rm H}$ is the horizon length. This result has been shown to hold in both radiation and matter eras.
As a consequence of this highly efficient mechanism of annihilation, the problem of monopole over-production during phase transitions in the early universe does not arise for global monopoles.

\subsection[$I^2 FF$ theories and minimal coupling]{$\mathbf{I^2 FF}$ theories and minimal coupling}
\label{sec:iff}

We now consider gauge theories of the
type~\cite{Ratra:1991bn} which breaks the Weyl conformal symmetry of the gauge sector through a modification of the kinetic sector of the vector boson field,
\begin{equation}
 \frac{\mathcal{L}}{\sqrt{-g}} \supset -\frac{I^2}{4} F_{\mu \nu} F^{\mu
  \nu} .
\label{I2FF}
\end{equation}
Here $F_{\mu\nu}$ is the field strength of the $U(1)$ (hypercharge or electromagnetism, depending on the energy scale) gauge field, and $I (>0)$ represents a generic coupling between the gauge field and some other fields.
Supposing that $I$ is constant in space, then it can be absorbed into the vector field by the redefinitions
\begin{equation}
\label{eq:redef}
    \tilde{A}^a_\mu = I A^a_\mu,~~~\tilde{e} = e / I .
\end{equation}
Under this redefinition, one can see that the effect of $I$ is merely to shift the gauge coupling~$e$, and in particular it is clear that a small~$I$ would lead to strong couplings in the electric sector~\cite{Demozzi:2009fu}. 
To avoid this problem, we assume $I \geq 1$ and restrict our attention to the case of a monotonically decreasing function until it reaches unity.
As a toy model, we assume that the time evolution of the function $I$ can be expressed as
\begin{equation}
  I = 
 \begin{dcases}
     \left(\frac{a_{\mathrm{con}}}{a}\right)^{s} 
       & \text{for $a \leq a_{\mathrm{con}}$,} \\
     1
       & \text{for $a \geq a_{\mathrm{con}}$,}
 \end{dcases}
 \label{power-law}
\end{equation}
where $s$ is a positive integer. 
Here we define $t_{\rm con}$ as the time when the model recovers the conformal invariance of the gauge sector, i.e. $I(t_{\rm con}) = I_{\rm con} = 1$ (the subscript ``con'' denotes quantities computed at $t_{\rm con}$). We also assume the temperature $T_{\rm con}$ to be larger than $1~\mathrm{MeV}$ to avoid constraints from Big Bang Nucleosynthesis (BBN).

We now want to embed the conformal-violating term in Eq.~\eqref{I2FF} into the toy model of a $G = $ SO(3) gauge theory with a triplet Higgs~\cite{tHooft:1974kcl,Polyakov:1974ek}. 
Here we focus on the case of a minimal coupling with $I$, while the scenario of dilatonic coupling is discussed in Appendix~\ref{sec:dilatonic}.
In the minimal coupling case, the Weyl symmetry-violating coupling~$I$ appears only in the gauge kinetic term,
\begin{equation}
\label{eq:min_coup}
    \frac{\mathcal{L}}{\sqrt{-g}}  = -\frac{I^2}{4} F^a_{\mu \nu} F^{a \mu \nu} - \frac{1}{2} (D_\mu \phi)^a (D_\mu \phi)^a - \frac{\lambda}{8} (\phi^a \phi^a - v^2)^2 ,
\end{equation}
where $a = 1,2,3$, $F^a_{\mu \nu} = \partial_\mu A^a_\nu - \partial_\nu A^a_\mu + e \varepsilon^{abc} A^b_\mu A^c_\nu $, $(D_\mu \phi)^a = \partial_\mu \phi^a + e \varepsilon^{abc} A^b_\mu \phi^c$, $\varepsilon^{abc}$ is the Levi-Civita symbol, and $\lambda, v > 0$. 
We therefore canonically redefine the gauge field and the gauge coupling as in Eq.~\eqref{eq:redef}.

After the SO(3) symmetry breaks at the critical temperature~$T_c \sim v$,
the particle spectrum consists of a Higgs boson with mass $M_{\ro{H}} = \sqrt{\lambda} v$, one massless gauge boson, and two massive gauge bosons with mass $M_{\ro{V}} = \abs{\tilde{e}} v$.
In addition, the theory admits topological monopole solutions carrying fundamental magnetic charge $\tilde{g} = \pm 4\pi / \tilde{e}$, with mass $M_{\ro{M}} \sim \abs{\tilde{g}} v = 4 \pi v / \abs{\tilde{e}}$, and size $r_{\ro{M}} \sim 1 / M_{\ro{V}} = 1 / \left(\abs{\tilde{e}} v \right)$.\footnote{Monopole solutions with higher magnetic charge are also possible, with charge given by integer multiples of the fundamental value: $\tilde{g} =  4 n \pi / \tilde{e}$, where $n$ is an integer.}
The redefinition of the gauge coupling leads to the following $I$-dependence of the gauge boson mass, as well as of the monopole’s charge, mass, and radius,
\begin{equation}
 M_{\ro{V}} \propto I^{-1}, \, 
 \tilde{g} \propto I, \, 
 M_{\ro{M}} \propto I, \, 
 r_{\ro{M}} \propto I.
\label{eq:scaling-1}
\end{equation}
Since $I$ decreases with time, a large value of $I$ in the asymptotic past corresponds to a strong coupling for the monopoles, i.e. a large~$\tilde{g}$.
Moreover, one finds that $M_{\ro{V}} / v$ decreases towards the past.
Thus, in the asymptotic past the condition $M_{\ro{V}} < H < T_c \sim v$ may be satisfied.
In this regime, the monopoles behave effectively as global objects, meaning they are chargeless under the gauge group, as only the global part of the symmetry is effectively spontaneously broken (the boson vector fields are still effectively massless, and the magnetic monopole radius extends beyond the Hubble horizon).
When $I$ approaches unity, the theory eventually enters the regime
$H < M_{\ro{V}} < T_c \sim v$, in which the monopoles become dressed by the gauge fields and acquire magnetic charge.\footnote{If, at the time monopoles become local, the magnetic charge $\tilde{g}$ remains sufficiently large, strong primordial magnetic fields could lead to efficient monopole–antimonopole pair production through the magnetic dual of the Schwinger effect \cite{Schwinger:1951nm,Affleck:1981ag,Affleck:1981bma,Kobayashi:2021des}.
This may act as an alternative production mechanism for monopoles.}
We will discuss the implications of these considerations for the monopole abundance in the next section.

\section{Solution of the monopole problem without inflation}
\label{sec:solution}

The standard computation of monopole abundance in the absence of modifications to the gauge sector is presented in Appendix~\ref{sec:phase_transition}. Based on those results, the monopole problem can be reformulated as follows: monopoles produced after inflation during a phase transition with critical temperature $T_c \gtrsim 10^{11}~\mathrm{GeV}$ would overclose the Universe.
In this section, we show that in the model defined by Eq.~\eqref{eq:min_coup}, it is possible to avoid the monopole problem, even for large vacuum expectation values.

For simplicity, we assume that radiation domination holds throughout the period considered for this analysis. We also assume that the Universe remains thermalized and conductive during this time, so that effects from post-inflationary magnetogenesis induced by the time variation of the $I$ function \cite{Ratra:1991bn,Demozzi:2009fu,Kobayashi:2014sga,Kobayashi:2019uqs} can be neglected \cite{Durrer:2013pga}. 
While more accurate consideration would be necessary to support this assumption, this would require the study of magnetohydrodynamic turbulence and lies beyond the scope of this work. We therefore leave it for future studies.\footnote{One should also notice that, in a post-inflationary scenario, some additional sector responsible for conformal symmetry breaking would be required to transfer energy into the gauge fields.}
We also note that, in general, the time dependence of the $I$ function can lead to efficient particle production. However, any relativistic particles produced before BBN and in equilibrium with the thermal bath remain phenomenologically irrelevant, while the production of massive particles is exponentially suppressed. Therefore, we neglect any possible contribution from particle production.\footnote{Our model may still generate potentially observable signatures, such as a gravitational-wave background. We leave this for future studies.}

When the temperature of the Universe falls below the critical temperature of the phase transition, $T \lesssim v$, and the Hubble rate is still larger than the vector boson mass, $H \gg \tilde{e} v = e v / I$, the gauge sector remains decoupled from the scalar sector and monopoles are produced as global. We denote by $T_{\mathrm{global}} = v$ the temperature at which global monopole production occurs (the subscript ``global'' corresponds to quantities computed at the time of the global symmetry breaking $t_{\rm global}$).
Later, when the Hubble rate becomes comparable to the vector boson mass, the local gauge part of the symmetry also breaks, and the monopoles acquire magnetic charge. We define
\begin{align}
    H_{\mathrm{local}} & = \frac{e}{I_{\mathrm{local}}} v , \\
    T_{\mathrm{local}}^2 & = \frac{\sqrt{90}}{\pi}~g_{*,\mathrm{local}}^{-1/2}~\frac{e}{I_{\mathrm{local}}} M_{\rm pl} v \sim \frac{e}{I_{\mathrm{local}}} M_{\rm pl} v,
\end{align}
as the Hubble rate and the temperature at the time of local symmetry breaking.
Here we denote with the subscript ``local'' denotes quantities computed at the time of the local symmetry breaking $t_{\rm local}$ and we use the equality $3 \Mp^2 H^2 = \left( \pi^2/30 \right) g_* T^4$ for the second equation. Hereafter, we also fix for simplicity $g_{*(s)} = 10$ throughout the radiation dominated epoch before BBN.
During the interval $T_{\rm global} > T > T_{\rm local}$, monopoles behave as global objects, and therefore their number density is fixed to the value given in Eq.~\eqref{eq:nGlobal}, meaning that the monopole number per Hubble volume is approximately conserved.
Once the temperature drops below $T_{\rm local}$, local symmetry breaking takes place and the monopoles transition to their standard evolution, and the number density scales as for non-relativistic matter.
The global-to-local transition, driven by the time evolution of the $I(t)$ rather than the Hubble rate, occurs on a timescale $|\dot{I}/I|^{-1} \sim H^{-1}$, comparable to the Hubble time. 
The effect of this transition can be interpreted as the effective radius of the gauge interaction shrinking over roughly one Hubble time -- from being larger than the Hubble radius, where defects behave as global and interactions are long-ranged, to being much smaller. The topology of the monopoles in the scalar field configuration remains unchanged, while the gauge-field monopole solution only emerges after the transition, ``dressing'' the defect with a magnetic charge. 

\begin{figure}[!t]
  \centering
  \includegraphics[width=\textwidth]{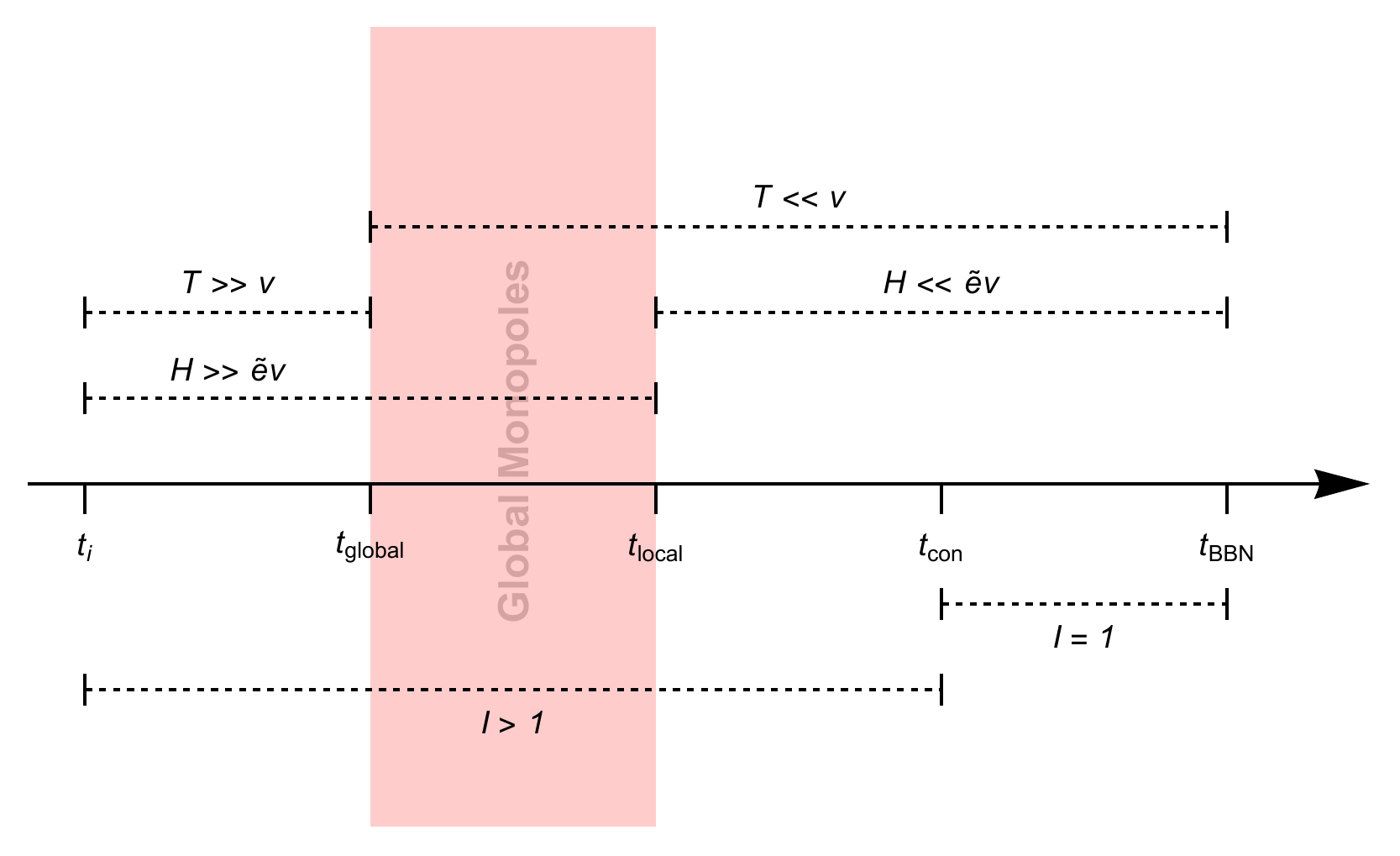}
  \caption{Timeline of the model considered in this work. We denote by $t_{\rm i}$ the onset of radiation domination, $t_{\rm global}$ the time of global symmetry breaking, $t_{\rm local}$ the time of local symmetry breaking, $t_{\rm con}$ the restoration of conformal symmetry in the gauge sector, and $t_{\rm BBN}$ the epoch of BBN. The red region evidences the time interval during which the monopoles act as global topological defects.}
  \label{fig:timeline}
\end{figure}
Figure~\ref{fig:timeline} illustrates the timeline of our model. For $t > t_{\rm i}$ the Universe is radiation dominated. At $t = t_{\rm global}$ the global symmetry breaks, and monopoles are produced during the Phase transition as global topological defects. In the interval $t_{\rm global} < t < t_{\rm local}$ ($H \gg \tilde{e} v,~T \ll v$), these monopoles follow the scaling solution given in Eq.~\eqref{eq:nGlobal}. In the figure, we highlight the time interval during which the monopoles are global with the red region. 
At $t = t_{\rm local}$ the local part of the symmetry also breaks and the monopoles acquire magnetic charge. Therefore, for $t > t_{\rm local}$ ($H \ll \tilde{e} v,~T \ll v$) they redshift as non-relativistic matter. Finally, at $t = t_{\rm con}$ the conformal symmetry of the gauge sector is restored ($I = 1$), and $t_{\rm con} < t_{\rm BBN}$ ensures our model does not affect BBN.

Considering that a monopole solution can be defined only for $T \lesssim v$, we do not lose generality by assuming non-relativistic monopoles.
Therefore, we define the monopole energy density as that for non-relativistic monopoles, 
\begin{equation}
    \rho_{\rm M} = n_{\rm M} M_{\ro{M}} .
\end{equation}
Notice that in our model the monopole mass $M_{\ro{M}}$ is time-dependent for $t<t_{\rm con}$. 
Taking from Eq.~\eqref{eq:nGlobal} the estimate for the number density $n_{\rm M, local} \approx 4 \times H_{\rm local}^3$ and considering that non-relativistic monopoles redshift as matter, the present-day energy density can be written as
\begin{equation}
\label{eq:rho_global}
    \rho_{\mathrm{M},\mathrm{glo}} = n_{\mathrm{M, local}}\ m_{\rm M} \left( \frac{a_{\mathrm{local}}}{a_0} \right)^3 \sim \rho_{\rm crit} \left( \frac{v}{10^{11}~\mathrm{GeV}} \right) \left( \frac{T_{\mathrm{local}}}{10^{11}~\mathrm{GeV}} \right)^3 ,
\end{equation}
where $m_{\rm M} = 4 \pi v / e$ is the monopole mass after the restoration of conformal symmetry and $\rho_{\rm crit} \approx 3.7 \times 10^{-47}~\mathrm{GeV}^4$ is the critical density of the universe today.
Let us compare this result with the standard estimate of the monopole energy density obtained via the Kibble mechanism
\cite{Kibble:1976sj},\footnote{More recent studies (see for instance~\cite{Brummer:2025inh}) have shown that the monopole abundance predicted by the Kibble mechanism can increase by several orders of magnitude when adopting less conservative estimates for the Higgs field coherence length. Nevertheless, the standard estimate reported here is sufficient for the purposes of this work.}
\begin{equation}
\label{eq:rho_local}
    \rho_{\mathrm{M},\mathrm{loc}} = \rho_{\rm crit} \times \Omega_M h^2 \sim \rho_{\rm crit} \left( \frac{v}{10^{11}~\mathrm{GeV}} \right)^4 ,
\end{equation}
where $\Omega_M$ is computed in Eq.~\eqref{eq:omegaKibble}.
The result in Eq.~\eqref{eq:rho_global} shows that, for $T_{\rm local} \ll v$, our model estimates a smaller monopole energy density today, compared to the standard scenario.
Hence, for sufficiently low values of $T_{\rm local}$, much larger symmetry-breaking scales (even larger than the GUT scale) are compatible with cosmology, without leading to a monopole overabundance.

We now discuss the effective range of applicability of our model. The model has three degrees of freedom. We choose as free parameters the vacuum expectation value, $v$, the temperature at which the conformal symmetry is restored, $T_{\rm con}$, and the exponent $s$ for the time dependence of $I$ defined in Eq.~\eqref{power-law}.
We can therefore express $T_{\rm local}$ in terms of the chosen parameters as
\begin{equation}
    T_{\rm local} \sim \left(e \Mp v T_{\rm con}^s \right)^{\frac{1}{2+s}} .
\end{equation}

Our new estimate of the relic monopole energy density in Eq.~\eqref{eq:rho_global} is valid under the following assumptions:
$T_{\rm con} \gtrsim T_{\rm BBN} \sim 1~\mathrm{MeV}$, i.e. conformal symmetry must be restored before BBN in order not to disrupt standard cosmology;
$T_{\rm local} \gtrsim T_{\rm con}$, i.e. the local phase transition must occur prior to the restoration of conformal symmetry;
$T_{\rm global} > T_{\rm local}$, i.e. global symmetry breaking must precede the local one.
The last two conditions yield relations among the free parameters of the model $(v, T_{\rm con}, s)$.
In particular, the second condition implies
\begin{equation}
\label{eq:loc_con}
    T_{\rm con}^2 \gtrsim e M_{\rm pl} v ,
\end{equation}
while the third condition gives
\begin{equation}
\label{eq:glo_loc}
    v^{s+1} \gtrsim e \Mp T_{\rm con}^s .
\end{equation}
For realistic values of the parameters, the condition in Eq.~\eqref{eq:loc_con} is usually weaker than that in Eq.~\eqref{eq:glo_loc}, and therefore in the following we focus mainly on the condition in Eq.~\eqref{eq:glo_loc}.
If the condition in Eq.~\eqref{eq:glo_loc} is not verified, the symmetry breaking follows the usual pattern and occurs at the temperature $T \sim v$. Hence, the monopoles are produced directly as local and the energy density follows the expression in Eq.~\eqref{eq:rho_local}. 

\begin{figure}[!t]
  \centering
  \begin{subfigure}[b]{0.49\textwidth}
    \centering
    \includegraphics[width=\textwidth]{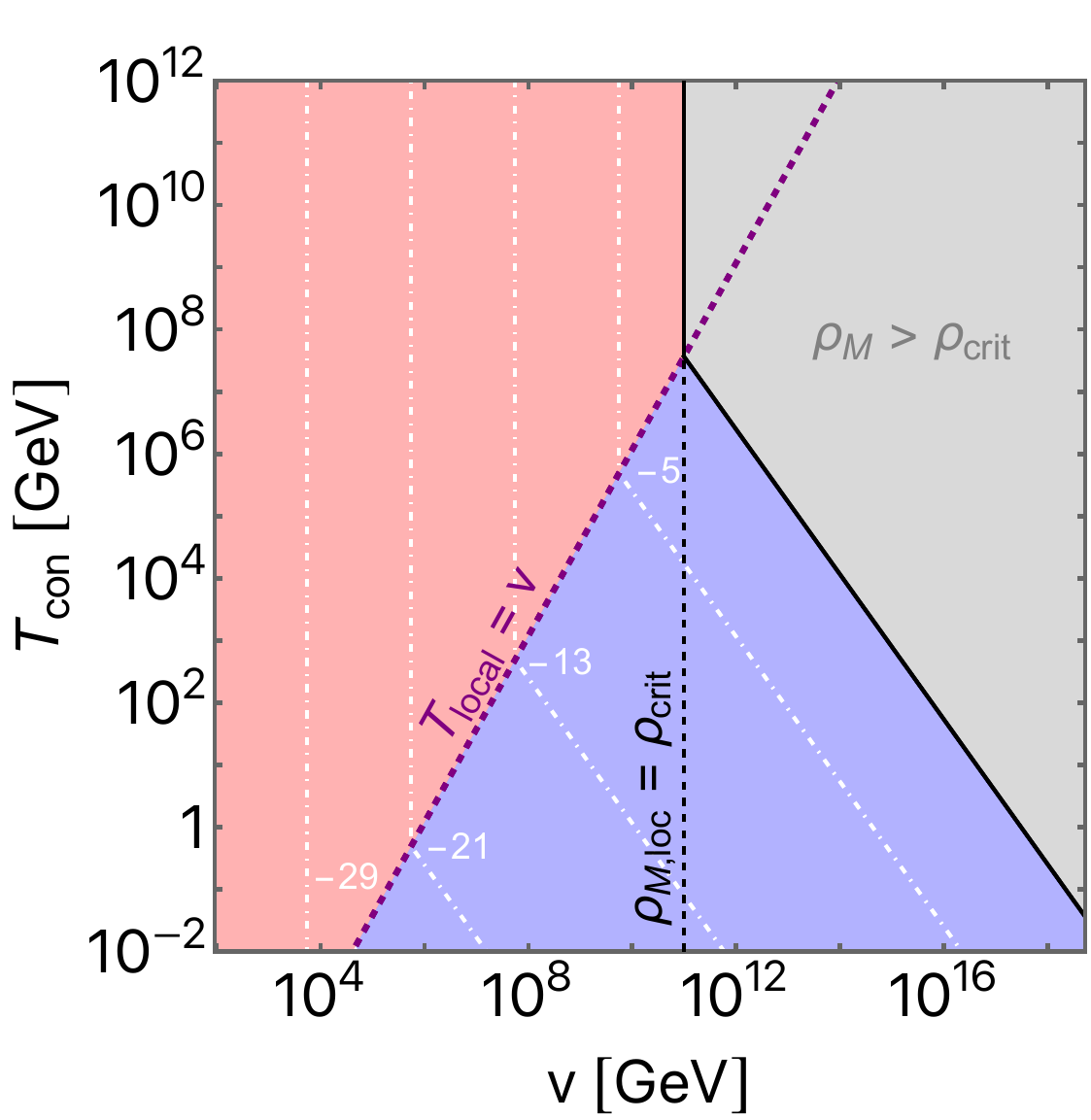}
    \caption{$s = 2$.}
    \label{fig:global3}
  \end{subfigure}
  \hfill
  \begin{subfigure}[b]{0.49\textwidth}
    \centering
    \includegraphics[width=\textwidth]{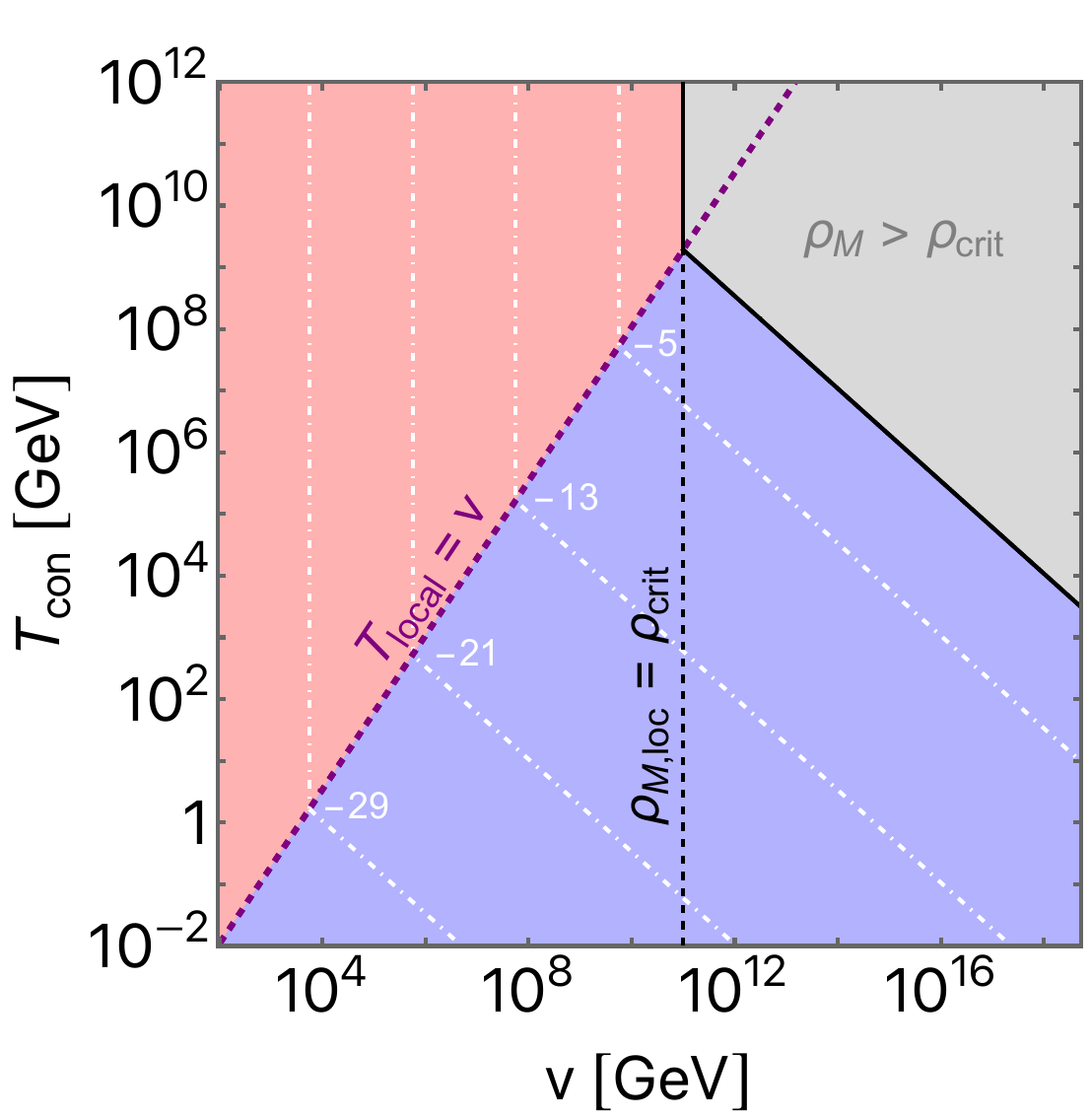}
    \caption{$s = 4$.}
    \label{fig:global6}
  \end{subfigure}
  \caption{Parameter space of the relic monopole energy density as a function of the vacuum expectation value $v$ and the temperature $T_{\rm con}$ at which conformal symmetry in the gauge kinetic sector is restored. Results are shown for two different values of the positive integer $s$ defined in Eq.~\eqref{power-law} ((a): $s=2$, (b): $s=4$). The gray region is excluded, as monopoles would overclose the Universe ($\rho_{\rm M} > \rho_{\rm crit}$). In the red region, monopoles are produced directly as local and their abundance is determined by Eq.~\eqref{eq:rho_local}. In the blue region, monopoles are initially produced as global and their abundance is given by Eq.~\eqref{eq:rho_global}. White dashed contours indicate the expected present-day monopole energy density in units of the critical density. The solid black line marks the parameter combinations for which the monopole density equals the critical density, while the solid black line marks those for which only the standard computation in Eq.~\eqref{eq:rho_local} equals the critical density. The solid purple line corresponds to $T_{\rm local} = v$. Here, before BBN, we assume $g_{*(s)} = 10$.
}
  \label{fig:global}
\end{figure}
In Figure~\ref{fig:global}, we show the monopole energy density as a function of the symmetry breaking scale $v$ and the temperature at the restoration of the conformal symmetry, $T_{\rm con}$. 
We show the results for two different values of the integer $s$ (Figure~\ref{fig:global3}: $s=2$, Figure~\ref{fig:global6}: $s=4$). In the plots, the limit of the axis are set in order to respect the conditions $T_{\rm con} \gtrsim T_{\rm BBN}$, and $\Mp > v > v_{\rm EW}$.\footnote{In an inflationary universe, values of $T_{\rm global} \sim v \gg 10^{16}~\mathrm{GeV}$ are excluded by upper bounds on the inflationary scale. In the plot, we also show larger values of the vacuum expectation value, since our main focus is on the implications of our model for resolving the monopole problem without inflation.} We now explain the plots in detail.

The purple line corresponds to parameter values for which the temperature of local symmetry breaking, $T_{\rm local}$, equals that of global symmetry breaking, $T_{\rm global}$, that is, when the equality in Eq.~\eqref{eq:glo_loc} holds. To the right of the purple line, global symmetry breaking occurs before the local one, so monopoles are initially produced as global. In this case the today monopole energy density is set by Eq.~\eqref{eq:rho_global}. Conversely, to the left of the purple line, both the gauge and the Higgs sector are affected by the symmetry breaking simultaneously, monopoles are produced directly as local defects via the standard Kibble mechanism, and the today monopole energy density is set by Eq.~\eqref{eq:rho_local}.
The solid black line indicates the parameter combinations for which in our model the monopole energy density equals the critical density of the Universe. Above this line, monopoles would overclose the Universe, while below it the abundance is cosmologically viable. 
The dashed vertical black line denotes the value of $v$ at which the standard estimate of the monopole energy density in Eq.~\eqref{eq:rho_local} equals the critical density. To the left of this line, monopoles pose no problem in the standard scenario, whereas to the right, standard estimates exclude monopole production after inflation. 
We show the region of the plot where in our model the monopole energy density exceeds the Universe in gray. 
Conversely, in the red and blue regions the monopole energy density remains below the critical density. In the red region, monopoles are produced directly as local defects, while in the blue region, monopoles are initially produced as global and later acquire magnetic charge.
The portion of the blue region to the right of the dashed black line corresponds to parameter values that are allowed in our model but excluded in the standard scenario.
The dashed white contours indicate the present-day ratio of the monopole energy density to the critical density of the Universe.

Our results show that modifying the kinetic term of the gauge sector as in Eq.~\eqref{I2FF} significantly broadens the range of admissible symmetry-breaking scales, allowing for larger vacuum expectation values. In particular, within our simplified setup, if the $I$ function exhibits a sufficiently strong redshift dependence ($s \geq 2$), the monopole problem can be solved up to the Planck mass scale, without requiring a post–monopole production inflationary epoch. Consequently, even GUT-scale monopoles can be generated during post-inflationary phase transitions without conflicting with cosmological constraints.
Furthermore, our framework predicts that, depending on the parameters, the resulting monopole abundance can be large enough to be probed both by terrestrial experiments~\cite{Perri:2025qpg,PierreAuger:2016imq,IceCube:2021eye,ANTARES:2022zbr} and by indirect searches~\cite{Turner:1982ag,Parker:1970xv,Parker:1987,Kobayashi:2022qpl,Kobayashi:2023ryr}.
In particular, we notice that even GUT-scale monopoles can constitute a substantial fraction, or the entirety, of dark matter. Thus, our model provides an efficient mechanism for monopole production in the early universe, positioning them as viable dark matter candidates.\footnote{For monopoles to be viable dark matter candidates, one should also check that they can cluster with galaxies and are not accelerated to relativistic velocities by cosmic magnetic fields. We refer the reader to \cite{Kobayashi:2023ryr,Perri:2023ncd,Perri:2025qpg} for a detailed discussion of these conditions.}

In Appendix~\ref{sec:dilatonic}, we consider the alternative scenario of a dilatonic coupling of $I^2$ to the Standard Model Lagrangian, and show how, differently from the minimal coupling discussed here, this alternative framework fails to resolve the monopole problem.

\section{Conclusion}
\label{sec:conclusion}

In this work, we Have proposed a new solution to the monopole overabundance problem arising from primordial phase transitions after inflation. Our model introduces a modification of the gauge sector kinetic term, breaking conformal invariance through a time-dependent coupling.
We have shown that within this framework monopoles are initially produced as global topological defects, leading to a substantial suppression of their abundance due to efficient monopole–antimonopole annihilation. This mechanism permits symmetry-breaking scales up to the Planck mass without encountering the monopole problem, thereby reopening the possibility of producing GUT-scale magnetic monopoles after inflation.

Our model provides an viable mechanism for the production of magnetic monopoles as cosmological relics. Unlike in standard scenarios, in our model the monopole abundance is not diluted by inflationary expansion, leaving a relic density potentially large enough to be accessible to both direct and indirect searches. These results highlight the importance of dedicated experimental efforts to probe GUT-scale monopoles, which remain only weakly constrained and therefore continue to stand as viable dark matter candidates.

\appendix

\section{Monopole production during phase transition in the early universe}
\label{sec:phase_transition}

The correlation length $\xi$ of the Higgs field sets the maximum distance over which the field remains correlated.
Due to this finite correlation length, if allowed by the theory, non-trivial vacuum configurations such as magnetic monopoles will inevitably be produced during phase transitions, with an approximate abundance of one per coherence volume $\xi^3$. The corresponding monopole number density at the transition can thus be estimated as $n_M \sim \xi^{-3}$.
As the correlation length depends on the specifics of the phase transition, its precise value might be difficult to determine.
In this appendix, we review the most commonly adopted estimates of the monopole abundance produced during a phase transition in the early universe.

In the case of second-order (or weakly first-order) phase transitions, the cosmological production of monopoles can be estimated through the ``Kibble mechanism'' \cite{Kibble:1976sj}, 
which closely resembles the method for generating topological defects in standard laboratory phase transitions. 
The key idea of the Kibble mechanism is that during a cosmological phase transition, the correlation length of the Higgs field cannot exceed the particle horizon, which is the furthest distance over which a massless particle could have traveled since the Big Bang, given by
\begin{equation}
    d_H = a(t) \int_{0}^{t} \frac{dt'}{a(t')} ,
\end{equation}
where $a(t)$ is the scale factor of the Universe. Hence, the horizon provides a strict upper limit on the correlation length and defines the natural scale used in the Kibble mechanism to estimate the Higgs field correlation length.

Following standard cosmology, the age of the Universe when the phase transition occurs ($t = t_c,~T = T_c$) is approximately given by $t_c \sim 0.3~g_*^{-1/2} \Mp/T_c^2$. 
According to the Kibble mechanism, magnetic monopoles are estimated to form with an approximate abundance of one monopole per horizon volume, corresponding to a number density of order $n_M \sim d_H^{-3}$.
Taking $d_H \sim t_c$, and recalling that the entropy density at temperature $T_c$ is $s \sim 2.3~g_{*s} T_c^3$, the monopole-to-entropy ratio is therefore given by
\begin{equation}
\label{eq:nms_first}
    \frac{n_M}{s} \sim 10^2~ \frac{g_*^{3/2}}{g_{*s}} \left( \frac{T_c}{\Mp} \right)^3 .
\end{equation}
Assuming negligible monopole-antimonopole annihilation\footnote{Preskill in \cite{Preskill:1979zi} has shown that monopole-antimonopole annihilation is significant only if $n_M/s \gtrsim 10^{-10}$. In that case, the ratio is reduced by annihilation to $n_M/s \sim 10^{-10}$.} and no entropy production, this ratio stays constant and sets the current abundance of monopoles. Under this assumption, the monopole eneergy density in units of the critical density can be written as
\begin{equation}
\label{eq:omegaKibble}
    \Omega_M h^2 \sim 10^{11} \left( \frac{T_c}{10^{14}~\mathrm{GeV}} \right)^3 \left( \frac{m_M}{10^{16}~\mathrm{GeV}} \right) \sim \left(\frac{v}{10^{11}~\mathrm{GeV}}\right)^4 ,
\end{equation}
where $v$ is the vacuum expectation value of the theory and the last equality comes from $T_c \sim v$ and $m_M \sim 4 \pi v/ e$.
Therefore, the safety condition $\Omega_M < 1$ translates into an upper limit for the critical temperature of the phase transition, that is, $T_c \sim v \ll 10^{11}~\mathrm{GeV}$.
For GUT monopoles, one typically has $T_c \sim 10^{14}~\mathrm{GeV}$ and $m_M \sim 10^{16}~\mathrm{GeV}$. This would yield a present-day monopole mass density about $10^{11}$ times larger than the critical density of the universe, which is clearly impossible. This inconsistency is usually addressed as the ``monopole problem''.

In the case of strongly first-order phase transitions (that Is, the transition proceeds by bubble nucleation at the nucleation temperature $T_n \ll T_c$, when the nucleation rate becomes comparable to the Hubble rate), the estimate of the monopole abundance should be slightly modified. 
Inside each bubble, the Higgs field is coherent; whereas, in distinct bubbles, it has no correlation. Consequently, it is expected that roughly one monopole forms per bubble. 
Specifically, as the Universe cools down to temperature $T_n$, bubbles start to form, expand, and quickly occupy all space. If $r_b$ represents the average bubble size during this phase, the expected monopole density is $n_M \sim r_b^{-3}$. 
Once the bubbles merge the Universe warms up again, and the entropy density approaches $s \sim g_* T_c^3$. Therefore, the monopole-to-entropy ratio becomes $n_M/s \sim (g_* r_b^3 T_c^3)^{-1}$.
It has been shown in \cite{Guth:1982pn} that the bubble size can be approximately expressed as
\begin{equation}
    r_b \sim \left( \Mp / T_c^2 \right) \left( \log \left( \Mp^4 / T_c^4 \right) \right)^{-1},
\end{equation}
resulting in a relic monopole abundance of
\begin{equation}
    \frac{n_M}{s} \sim \left[ \left( \frac{T_c}{\Mp} \right) \log \left( \frac{\Mp^4}{T_c^4} \right) \right]^3 .
\end{equation}
Therefore, comparing with Eq.~\eqref{eq:nms_first}, one sees that with a first order phase transition the monopole problem is still present, and, for realistic values of the critical temperature ($T_c \lesssim 10^{16}~\mathrm{GeV}$), is even worse.

\section{$I^2 FF$ theories with dilatonic coupling}
\label{sec:dilatonic}

In this appendix, we consider theories of the
type~\cite{Ratra:1991bn} which breaks the  Weyl conformal symmetry of the gauge sector with a dilatonic symmetry-violating coupling $I$. As in the case of Section~\ref{sec:iff}, we embed the conformal-violating term into the toy model of a $G = $ SO(3) gauge theory with a triplet Higgs~\cite{tHooft:1974kcl,Polyakov:1974ek},
\begin{equation}
 \frac{\mathcal{L}}{\sqrt{-g}}  = I^2 \left\{ -\frac{1}{4} F^a_{\mu \nu} F^{a \mu \nu} -\frac{1}{2} (D_\mu \phi)^a (D_\mu \phi)^a- \frac{\lambda}{8} (\phi^a \phi^a - v^2)^2 \right\} ,
\label{eq:case2}
\end{equation}
where $a = 1,2,3$, $F^a_{\mu \nu} = \partial_\mu A^a_\nu - \partial_\nu A^a_\mu + e \varepsilon^{abc} A^b_\mu A^c_\nu $, $(D_\mu \phi)^a = \partial_\mu \phi^a + e \varepsilon^{abc} A^b_\mu \phi^c$, $\varepsilon^{abc}$ is the Levi--Civita symbol, and $\lambda, v > 0$.
In this case, $I$ is absorbed by the redefinitions 
\begin{equation}
    \tilde{A}^a_\mu = I A^a_\mu,~\tilde{e} = e / I,~\tilde{\phi}^a = I \phi^a,~\tilde{v} = I v,~\tilde{\lambda} = \lambda / I^2 ,
\end{equation}
and we adopt for $I$ the same definition of Eq.~\eqref{power-law}.
In this model, symmetry breaking occurs at the critical temperature $T_c \sim \tilde{v}$, yielding a particle spectrum that includes a Higgs boson with mass $M_{\ro{H}} = \sqrt{\tilde{\lambda}} \tilde{v} = \sqrt{\lambda} v$, one massless gauge boson, and two massive gauge bosons with mass $M_{\ro{V}} = \abs{\tilde{e}} \tilde{v} = \abs{e} v$.
In addition, 't Hooft–Polyakov monopoles exist with fundamental magnetic charge $\tilde{g} = \pm 4\pi / \tilde{e}$, mass $M_{\ro{M}} \sim \abs{g} \tilde{v} = 4\pi \tilde{v}/\abs{\tilde{e}}$, and size $r_{\ro{M}} \sim 1/M_{\ro{V}} = 1/(\abs{e} v)$.
The parameter shifts therefore induce, in the dilatonic case, the following $I$-dependence of the vacuum expectation value, the monopole charge, and the monopole mass:
\begin{equation}
 \tilde{v} \propto I,\, 
 \tilde{g} \propto I, \, 
 M_{\ro{M}} \propto I^2 .
\label{eq:scaling-2}
\end{equation}

We define $T_{\rm local}$ as the temperature at which the local symmetry is broken ($H_{\rm local} = \tilde{e} \tilde{v} = e v$), and $T_{\rm global}$ as the temperature at which the global symmetry is broken ($T_{\rm global} = v)$. Hence, in the dilatonic case the two temperatures can be expressed as
\begin{equation}
\label{eq:T_dilatonic_glo}
    T_{\mathrm{global}} \sim \tilde{v} = v I_{\rm global} ,
\end{equation}
\begin{equation}
\label{eq:T_dilatonic_loc}
    T_{\rm local} \sim \Mp^{1/2} H_{\rm local}^{1/2} = (e \Mp v)^{1/2} ,
\end{equation}
where we assume $g_{*(s)} = 10$ before BBN.
Differently from the case discussed in Section~\ref{sec:iff}, $T_{\rm local}$ depends only on the vacuum expectation value of the theory and the gauge coupling, while $T_{\mathrm{global}}$ depends also on the specific theory of the conformal symmetry breaking.

Assuming $T_{\rm global} > T_{\rm local}$, the expression for the monopole energy density is the same of Eq.~\eqref{eq:rho_global}, with $T_{\rm local}$ now determined by Eq.~\eqref{eq:T_dilatonic_loc}. This gives
\begin{equation}
    \rho_{\mathrm{M},\mathrm{glo}} \sim \rho_{\rm crit} \left( \frac{e \Mp}{10^{11}~\mathrm{GeV}} \right)^{3/2} \left( \frac{v}{10^{11}~\mathrm{GeV}} \right)^{5/2} ,
\end{equation}
Comparing this with the standard expression for the monopole abundance in Eq.~\eqref{eq:rho_local}, we find that suppressing today’s monopole density via global monopole annihilation would require An extremely large vacuum expectation value,
\begin{equation}
    v \gg e \Mp ,
\end{equation}
which is nonphysical.
We therefore conclude that a dilatonic Weyl symmetry–violating coupling cannot resolve the cosmological monopole problem.

\acknowledgments
D.P. thanks Takeshi Kobayashi for his valuable advice and fruitful discussions.
This study was partially supported by the National Science Centre, Poland, under research grant no. 2020/38/E/ST2/00243.

\bibliographystyle{unsrt}
\bibliography{global}

\end{document}